# Finite volume simulation of arc: pinching arc plasma by high-frequency alternating longitudinal magnetic field

Xiaoliang Wang[1,2]
1 School of Physical Sciences, University of Science and Technology of China, Hefei 230026, China
2 College of Life Sciences, Zhejiang University, Hangzhou 310058, China
Correspondence: (X.W.) wxliang@mail.ustc.edu.cn;

## Abstract

Arc plasmas have promising applications in many fields. To explore their property is of interest. This paper presents detailed pressure-based finite volume simulation of argon arc. In the modeling, the whole cathode region is coupled to electromagnetic calculations to promise the free change of current density at cathode surface. In numerical solutions, the upwind difference scheme is chosen to promise the transport property of convective terms, and the SIMPLE (Semi-Implicit Method for Pressure Linked Equations) algorithm is used to solve thermal pressure. By simulations of the free-burning argon arc, the model shows good agreement with experiment. We observe an interesting phenomenon that argon arc concentrates intensively in the high-frequency alternating longitudinal magnetic field. Different from existing constricting mechanisms, here arc achieves to be pinched through a continuous transition between shrinking and expansion. The underlying mechanism is that via collaborating with arc's motion inertia, the applied high-frequency alternating magnetic field is able to effectively play a "plasma trap" role, which leads the arc plasma to be imprisoned into a narrower space. This may provide a new approach to constrict arc.

## 1. Introduction

In the welding field, how to improve the welding quality (such as reducing the welding width and increasing the welding depth) and cut the production cost is much concerned. The direct and effective solution is to reduce the welding area and as well as to elevate the energy flux on the surface of the workpiece. For example, the laser welding is massively used in precision machining, due to its extremely high energy flux (>$10^8$ W/$m^2$) and also very fine focusing performance. As the traditional welding technique, arc welding is widely applied to industrial fields, such as machining, metallurgy, material processing, chemical production and even environmental protection, due to arc plasmas' high temperature, high enthalpy, chemical activity and also low cost. Exploring arc's property is thus meaningful and necessary not only for scientific advance but also for practical applications. However, owing to the poor welding quality, such as the broad welding seam and shallow welding depth, the application of arc welding is much limited. Therefore, in the arc welding, how to constrict arc as soon as possible has been of interest to researchers.

TIG (tungsten inert gas) arc is one promising welding arc. Many methods have been proposed to constrict TIG arc. Direct methods are to increase the welding current [1,2] and to conduct the mechanical and fluid cooling compression. Some researchers obtain constricted arc by amplifying the ambient pressure [3,4], using the laser-arc hybrid welding [5,6] and smearing active fluxes on the anode [7]. Particularly, as early as 1980s, Cook and Eassa [8] had found that using the high-frequency pulsed welding current can also make arc constricted intensively. This method was further confirmed by the recent work [9,10].

Since arc is the partially ionized gas, the applied magnetic field thereby provides a new approach to control arc. Under the specific magnetic field configuration, arc plasma can be confined and make changes in its shape. Plenty of work has been performed on the magnetically controlled arc. For instance, some researchers utilized the permanent cusp magnetic field to clamp arc [11-13]. Zhainakov et al. [14] and Ukita et al. [15] investigated the influence of transverse magnetic field. Yosuke et al. [16] experimentally observed the oscillation scale of a large-scale arc (arc length is about 40 mm) under the alternating transverse magnetic field. Under the constant longitudinal (axial) magnetic field, it is also found that arc gets contracted [17-19]. But when the magnetic field is strong, arc plasma easily gets dispersed near the anode and presents a hollow "bell" shape [20-22] and even totally scattered [23] due to the strong rotation of arc plasma.

Arc plasma is a complex system, involving mass, momentum, heat and electricity transport phenomena. Conventional theoretical analysis based on simple assumptions is quite hard to describe arc accurately. The experiment is also difficult to deeply understand mechanisms underlying many practical problems, as conducting the measurement of some physical quantities of arc is challenging. The numerical experiment provides a unique opportunity to acquire those quantities by simulating the arc plasma system.

Mostly, arc plasma can be treated as the electrically conductive fluid described by magnetohydrodynamics (MHD) equations. The finite volume method (FVM), which is able to strictly promise the conservativeness of governing equations during numerical solutions and has been widely used in computational fluid dynamics (CFD), is also taken to solve MHD equations in most of simulations of arc systems [1,12,14,21,22,24-30]. These simulations are based on home codes or commercial CFD software. However, the details of numerical discretization and solution using FVM are rarely shown. Here, we will introduce the detailed implementation of the FVM scheme in modeling of argon arc. In our modeling, the whole cathode region is coupled to the electromagnetic computation [21,22,28,29], which can make the current density distributed on the cathode surface solved automatically. This is more realistic, compared with giving some specific distribution of current density near the cathode tip [1,24-27]. Besides, we will also explore the constricting mechanism of arc in the applied magnetic field and try to provide a new approach to confine arc plasmas.

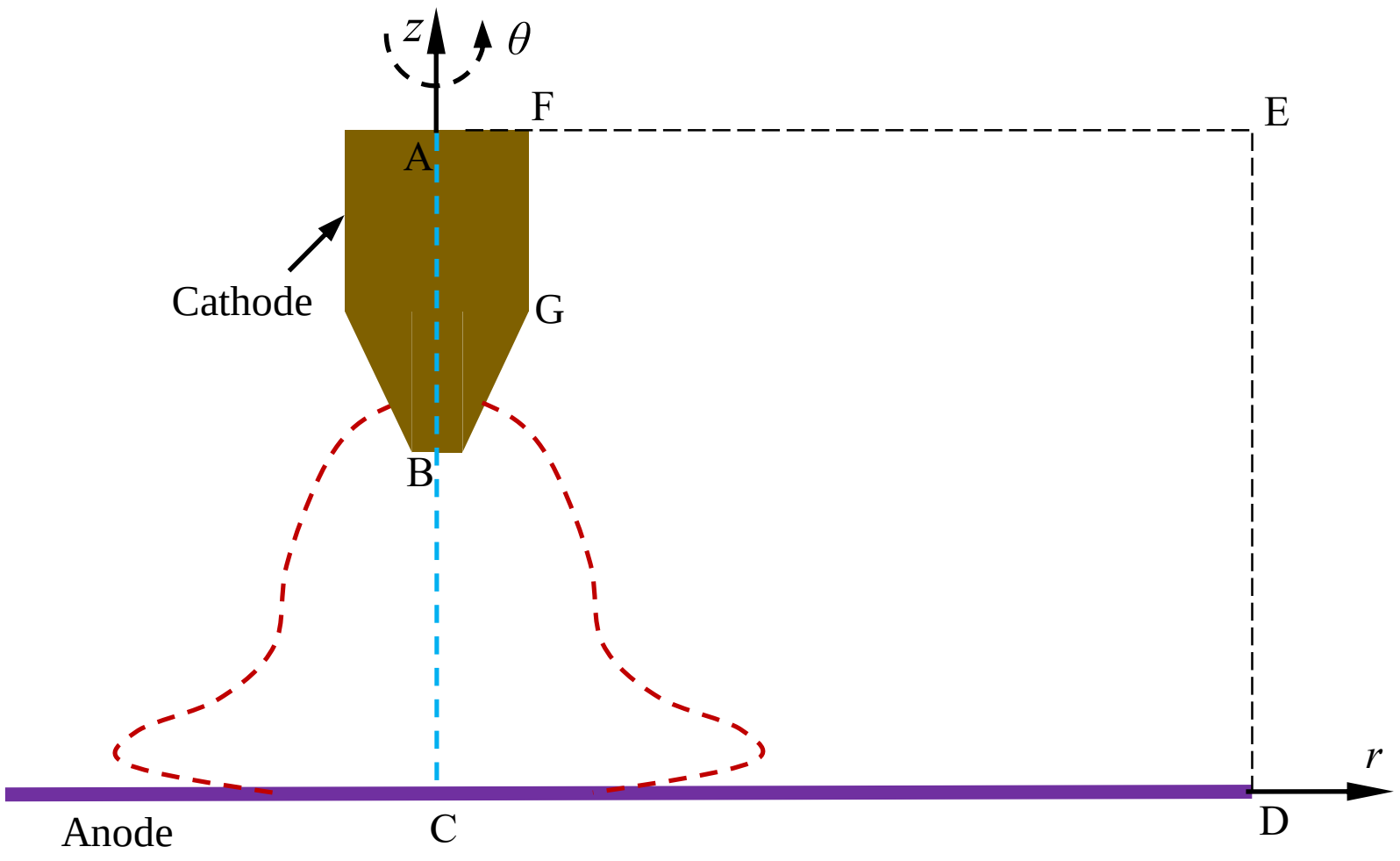

**Figure 1.** 2D illustration of the whole arc plasma region.

## 2. Models and methods

In this section, we present the detailed numerical solution of arc plasma, in terms of the steady free-burning argon arc. Shown in Fig. 1 is the whole arc plasma region selected for analysis. In the modeling, the cylindrical coordinate system is used and the main assumptions made for arc plasma are as follows:

- The arc is in local thermodynamic equilibrium (LTE) [1];
- The arc is steady and cylindrically symmetric;
- The gas flow in arc is laminar.

### 2.1 Governing equations

Based on assumptions above, the governing equations expressed in cylindrical coordinates ($z, r, \theta$) can be written as the following.

Mass conservation equation:

$$\frac{\partial \rho}{\partial t} + \nabla\cdot(\rho \boldsymbol{U}) = 0 \quad (1)$$

Momentum conservation equation:

$$\frac{\partial}{\partial t}(\rho \boldsymbol{U}) + \nabla\cdot(\rho \boldsymbol{U}\boldsymbol{U}) = -\nabla p + \nabla\cdot(\mu\nabla \boldsymbol{U}) + \boldsymbol{J}\times\boldsymbol{B} \quad (2)$$

Energy conservation equation:

$$\frac{\partial}{\partial t}\left(\rho c_p T\right) + \nabla\cdot\left(\rho c_p \boldsymbol{U} T\right) = \nabla\cdot(k\nabla T) + \frac{\boldsymbol{J}\cdot\boldsymbol{J}}{\sigma} + \frac{5}{2}\frac{k_B}{e}\cdot\boldsymbol{J}\nabla T - S_R \quad (3)$$

Current conservation equation:

$$\nabla\cdot\boldsymbol{J} = 0 \quad (4)$$

where $\boldsymbol{U} = (u,v,0)$ is velocity, and $u$ and $v$ represent the velocities in axial and radial directions, respectively. $p$ is the plasma pressure, $\boldsymbol{J} = (J_z,J_r,0)$ is current density, and $J_z$ and $J_r$ are respectively the axial and radial current density. $\boldsymbol{B} = (0,0,B_\theta)$ is magnetic field strength and $B_\theta$ is the self-induced magnetic field in the toroidal direction. $T$ is temperature. $\mu$, $k$ and $c_p$ are viscosity, thermal conductivity and specific heat, respectively. $k_B = 1.38\times10^{-23}$ J/K is the Boltzmann's constant, $e = 1.6\times10^{-19}$ C is the electron charge, and $S_R$ is the radiative source term.

To solve equations (1)-(4), some supplemental equations are needed and listed in the following:

Equation of state:

$$p = \rho R_g T \quad (5)$$

Ohm's law:

$$\boldsymbol{J} = \sigma \boldsymbol{E} \quad (6)$$

where $E = (E_z,E_r,0)$ is electric field strength and is expressed as:

$$\boldsymbol{E} = -\nabla\varphi \quad (7)$$

Ampere's law:

$$B_\theta = \frac{\mu_0}{r}\int_0^r J_z r'\,dr' \quad (8)$$

In Eqs. (5)-(8), $\varphi$ is the electrical potential in the arc, and $R_g$, $\sigma$ and $\mu_0$ are the gas constant, electrical conductivity and the permeability of vacuum, respectively. Electrical potential $\varphi$ can be obtained from the following Laplace's equation, which is derived from equations (4), (6) and (7).

$$\nabla\cdot(\sigma\nabla\varphi) = 0 \quad (9)$$

### 2.2 Boundary conditions

In the modeling, two domains, i.e. A-B-C-D-E-F-A and B-C-D-E-F-G-B are chosen. The big one including the whole cathode region is used only to calculate electromagnetic fields, while another one is used to solve the mass, momentum and energy conservation equations.

The centerline A-B-C is the axis of the arc system. On this boundary, the symmetry condition is employed to independent variables $u$, $v$, $p$, $T$ and $\varphi$. At the cathode surface B-G-F and anode surface C-D, a no-slip condition is postulated for flow velocities. In simulations, boundaries D-E and E-F can be chosen to be as far as possible from the arc plasma region so that the fully-developed assumption (the normal gradient at the boundary is zero, i.e. $\partial\phi/\partial n = 0$. $\phi$ is the general variable) and even the far-field condition (close to ambient conditions) can be used. In the whole arc region, A-F is the critical boundary, through which the current will flow to the tip to induce arc. At the boundary A-F, we use the uniform current density $J_0$, which is determined via dividing the total arc current by the cross sectional area of cathode. These boundary conditions are summarized in Table 1.

Note that due to the assumption of LTE, the temperature of electrons in the whole arc region is obliged to be equal to the heavy particles in calculations. LTE is farfetched for arc plasma's fringes, where the thermodynamic nonequilibrium would occur, leading the current continuity at plasma-electrode interfaces to be hard to promise while simulations. To handle this problem, we have adopted the solutions similar to those in [24,28,29]. One can visit these literatures for details.

Table 1: Boundary conditions.

| | $u$ | $v$ | $p$ | $T$ | $\varphi$ |
|---|---|---|---|---|---|
| A-B-C | $\partial u/\partial r = 0$ | 0 | $\partial p/\partial r = 0$ | $\partial T/\partial r = 0$ | $\partial \varphi/\partial r = 0$ |
| C-D | 0 | 0 | $\partial p/\partial n = 0$ | given | const. |
| D-E | $\partial u/\partial r = 0$ | $\partial v/\partial r = 0$ | 1 atm | $\partial T/\partial r = 0$ or fixed | $\partial \varphi/\partial r = 0$ |
| E-F | $\partial u/\partial z = 0$ | $\partial v/\partial z = 0$ | 1 atm | $\partial T/\partial z = 0$ or fixed | $\partial \varphi/\partial z = 0$ |
| F-A | - | - | - | - | $J_0$ |
| B-G-F | 0 | 0 | $\partial p/\partial n = 0$ | 3000 K | coupled |

### 2.3 Numerical discretization

Here, the FVM scheme is implemented to discretize conservation equations.

- Time discretization

The discretization of evolution equations in time can be implemented by means of the first-order Euler scheme, where the diffusion term is treated implicitly and the convection term is treated explicitly. In terms of the momentum equation, i.e. Eq. (2), we denote by $\boldsymbol{U}^n$ the approximation of $\boldsymbol{U}$ at time $t_n = n\Delta t$, where the super script $n$ is the natural number and $\Delta t$ is the time step length. Thus, the Euler semi-discretized form is

$$\frac{\rho^n \boldsymbol{U}^n}{\Delta t} - \nabla\cdot(\mu\nabla\boldsymbol{U}^n) + \nabla p^n = \frac{\rho^{n-1}\boldsymbol{U}^{n-1}}{\Delta t} - \nabla\cdot\left(\rho^{n-1}\boldsymbol{U}^{n-1}\boldsymbol{U}^{n-1}\right) + \boldsymbol{J}^{n-1}\times\boldsymbol{B}^{n-1} \quad (10)$$

Similarly, the Euler semi-discretized form for energy equation reads:

$$\frac{\rho^n c_p^n T^n}{\Delta t} - \nabla\cdot(k\nabla T^n) = \frac{\rho^{n-1}c_p^{n-1}T^{n-1}}{\Delta t} - \nabla\cdot\left(\rho^{n-1}c_p^{n-1}U^{n-1}T^{n-1}\right) + \frac{5}{2}\frac{k_B}{e}\cdot J^{n-1}\nabla T^{n-1} + \frac{J^{n-1}\cdot J^{n-1}}{\sigma} - S_R^{n-1} \quad (11)$$

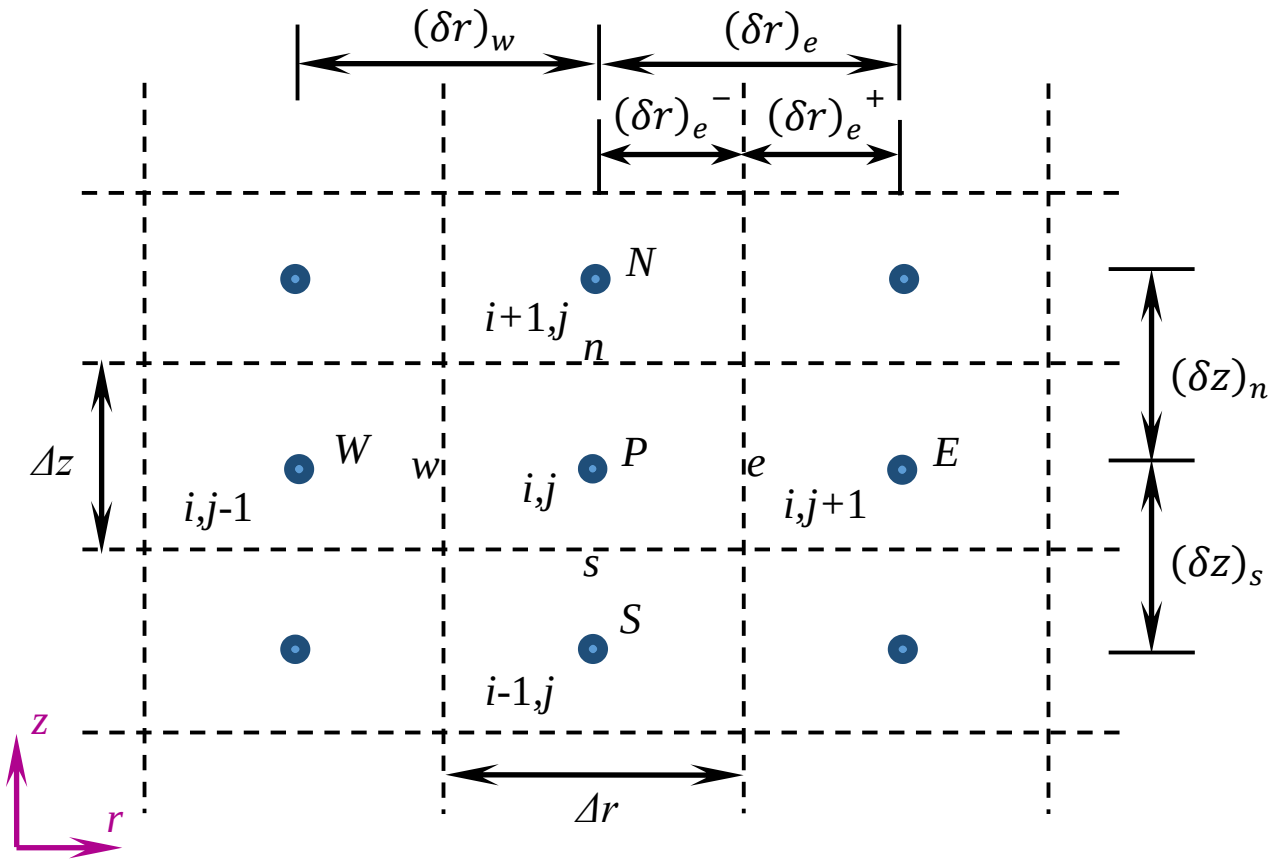

**Figure 2.** Schematic diagram of the structured grid in the cylindrical coordinate system.

- Space discretization

We chose the structured grid shown in Fig. 2 to present the basic idea of FVM. In Fig.2, the grid consisting of dashed lines produces a group of volumes around calculation nodes, and these volumes are next to each other. We take the component $u$ of Eq. (10) to show its discretization in space. The differential form of the component $u$ of Eq. (10) in space is expressed as:

$$\frac{\rho^n u^n}{\Delta t} - \left[\frac{\partial}{\partial z}\left(\mu\frac{\partial u^n}{\partial z}\right) + \frac{1}{r}\frac{\partial}{\partial r}\left(\mu r\frac{\partial u^n}{\partial r}\right)\right] + \frac{\partial p^n}{\partial z} = \frac{\rho^{n-1}u^{n-1}}{\Delta t} - \left[\frac{\partial}{\partial z}\left(\rho^{n-1}u^{n-1}u^{n-1}\right) + \frac{1}{r}\frac{\partial}{\partial r}\left(r\rho^{n-1}v^{n-1}u^{n-1}\right)\right] + J_r^{n-1}B_\theta^{n-1} \quad (12)$$

Integrate the Eq. (12) over the controlled volume $r\Delta z\Delta r$ represented by the grid node $P$ in Fig. 2, and we have,

$$\frac{\rho_{i,j}^n u_{i,j}^n}{\Delta t} r_P\Delta z\Delta r - \left[\left(r\mu\frac{\partial u^n}{\partial z}\right)_n \Delta r - \left(r\mu\frac{\partial u^n}{\partial z}\right)_s \Delta r + \left(r\mu\frac{\partial u^n}{\partial r}\right)_e \Delta z - \left(r\mu\frac{\partial u^n}{\partial r}\right)_w \Delta z\right] + r_P\Delta r\left(p_n^n - p_s^n\right) = \frac{\rho_{i,j}^{n-1}u_{i,j}^{n-1}}{\Delta t} r_P\Delta z\Delta r$$
$$- \left[(r\rho^{n-1}u^{n-1}u^{n-1})_n\Delta r - (r\rho^{n-1}u^{n-1}u)_s\Delta r + (r\rho^{n-1}v^{n-1}u^{n-1})_e\Delta z - (r\rho^{n-1}v^{n-1}u^{n-1})_w\Delta z\right] + \left(J_r^{n-1}B_\theta^{n-1}\right)_{i,j}\cdot r_P\Delta z\Delta r \quad (13)$$

$$\left(r\mu\frac{\partial u^n}{\partial z}\right)_n = r_n\mu_n\frac{u_{i+1,j}^n - u_{i,j}^n}{\Delta z},\ \left(r\mu\frac{\partial u^n}{\partial z}\right)_s = r_s\mu_s\frac{u_{i,j}^n - u_{i-1,j}^n}{\Delta z},\ \left(r\mu\frac{\partial u^n}{\partial r}\right)_e = r_e\mu_e\frac{u_{i,j+1}^n - u_{i,j}^n}{\Delta r},\ \left(r\mu\frac{\partial u^n}{\partial r}\right)_w = r_w\mu_w\frac{u_{i,j}^n - u_{i,j-1}^n}{\Delta r}$$

where the subscripts $s$, $n$, $e$ and $w$ represent the four boundaries of the controlled volume. The quantities distributed on these boundaries have been assumed to be uniform and can be evaluated by the linear interpolation from node values. The Eq. (13) conveys a clear physical meaning that in unit time, the total momentum increment within the controlled volume $P$ is provided by the net momentum that flows into and flows out through the interfaces of the volume $P$ and the forces acting upon the volume $P$, including the pressure, viscous resistance and Lorentz force.

Note that in the derivation from the Eq. (12) to the Eq. (13), the correct discretization form of the pressure gradient $-\partial p/\partial z$ is $-r_P\Delta r\left(p_n - p_s\right)$ in Eq. (13) instead of $-\Delta r\left(r_n p_n - r_s p_s\right)$ or some other forms. That is, the radius $r$ in front of both $p_n$ and $p_s$ should be $r_P$, since $-r_P\Delta r\left(p_n - p_s\right)$ represents the discretization for the pressure gradient, whereas $-\Delta r\left(r_n p_n - r_s p_s\right)$ is for the pressure divergence. If this detail is not noticed, the severe numerical error will occur. Generally, in Cartesian coordinates this problem won't happen since the radius $r$ doesn't exist there.

To promise the transport property of convective terms, here we chose the first-order upwind difference scheme, which is defined as follows:

$$u\frac{d\phi}{dx}\bigg|_i = \begin{cases} u_i\dfrac{\phi_i - \phi_{i-1}}{\Delta x}, & u_i > 0 \\ u_i\dfrac{\phi_{i+1} - \phi_i}{\Delta x}, & u_i < 0 \end{cases} \quad (14)$$

Note that the convective terms considered in this article are general. They include not only the mass convection through the interfaces of the controlled volume, but also the electricity convection appearing in the energy equation (the third term of the right hand of Eq. 3). The discretization of both mass and electricity convection terms is implemented with the upwind difference scheme.

Take the first-order upwind difference into the Eq. (13), and the right hand of Eq. (13) will become as:

$$S_u^{n-1} = \frac{\rho_{i,j}^{n-1}u_{i,j}^{n-1}}{\Delta t}r_P\Delta z\Delta r - \begin{bmatrix} \frac{r_n\Delta r}{2}\rho_n^{n-1}\left(u_n^{n-1}-\left|u_n^{n-1}\right|\right)\left(u_{i+1,j}^{n-1}-u_{i,j}^{n-1}\right)+\frac{r_s\Delta r}{2}\rho_s^{n-1}\left(u_s^{n-1}+\left|u_s^{n-1}\right|\right)\left(u_{i,j}^{n-1}-u_{i-1,j}^{n-1}\right) \\ +\frac{r_e\Delta z}{2}\rho_e^{n-1}\left(v_e^{n-1}-\left|v_e^{n-1}\right|\right)\left(u_{i,j+1}^{n-1}-u_{i,j}^{n-1}\right)+\frac{r_w\Delta z}{2}\rho_w^{n-1}\left(v_w^{n-1}+\left|v_w^{n-1}\right|\right)\left(u_{i,j}^{n-1}-u_{i,j-1}^{n-1}\right) \end{bmatrix} + \left(J_r^{n-1}B_\theta^{n-1}\right)_{i,j}\cdot r_P\Delta z\Delta r$$

where $|\cdot|$ denotes the absolute value symbol.

Differential equations for $v$, $T$ and $\varphi$ can be discretized in the same way. We can derive discretized equations for solving $u$, $v$, $T$ and $\varphi$ as the following.

The discretized equation for $u$:

$$a_S^u u_{i-1,j}^n + a_W^u u_{i,j-1}^n + a_P^u u_{i,j}^n + a_N^u u_{i+1,j}^n + a_E^u u_{i,j+1}^n + r_P\Delta r\left(p_n^n - p_s^n\right) = S_u^{n-1} \quad (15)$$

where $a_P^u = \frac{\rho_{i,j}^n}{\Delta t}r_P\Delta z\Delta r - (a_E^u+a_W^u+a_N^u+a_S^u)$,

$$a_E^u = - r_e\mu_e\frac{\Delta z}{\Delta r},\ a_W^u = - r_w\mu_w\frac{\Delta z}{\Delta r}$$

$a_N^u = - r_n\mu_n\frac{\Delta r}{\Delta z}$ and $a_S^u = - r_s\mu_s\frac{\Delta r}{\Delta z}$.

The discretized equation for $v$:

$$a_S^v v_{i-1,j}^n + a_W^v v_{i,j-1}^n + a_P^v v_{i,j}^n + a_N^v v_{i+1,j}^n + a_E^v v_{i,j+1}^n + r_P\Delta z\left(p_e - p_w\right) = S_v^{n-1} \quad (16)$$

where $a_E^v = a_E^u,\ a_W^v = a_W^u,\ a_N^v = a_N^u,\ a_S^v = a_S^u,\ a_P^v = a_P^u + \frac{\mu_P}{r_P}\Delta z\Delta r$ and

$$S_v^{n-1} = \frac{\rho_{i,j}^{n-1}v_{i,j}^{n-1}}{\Delta t}r_P\Delta z\Delta r - \begin{bmatrix} \frac{r_n\Delta r}{2}\rho_n^{n-1}\left(u_n^{n-1}-\left|u_n^{n-1}\right|\right)\left(v_{i+1,j}^{n-1}-v_{i,j}^{n-1}\right)+\frac{r_s\Delta r}{2}\rho_s^{n-1}\left(u_s^{n-1}+\left|u_s^{n-1}\right|\right)\left(v_{i,j}^{n-1}-v_{i-1,j}^{n-1}\right) \\ +\frac{r_e\Delta z}{2}\rho_e^{n-1}\left(v_e^{n-1}-\left|v_e^{n-1}\right|\right)\left(v_{i,j+1}^{n-1}-v_{i,j}^{n-1}\right)+\frac{r_w\Delta z}{2}\rho_w^{n-1}\left(v_w^{n-1}+\left|v_w^{n-1}\right|\right)\left(v_{i,j}^{n-1}-v_{i,j-1}^{n-1}\right) \end{bmatrix} - \left(J_z^{n-1}B_\theta^{n-1}\right)_{i,j}\cdot r_P\Delta z\Delta r.$$

The discretized equation for $T$:

$$a_S^T T_{i-1,j}^n + a_W^T T_{i,j-1}^n + a_P^T T_{i,j}^n + a_N^T T_{i+1,j}^n + a_E^T T_{i,j+1}^n = S_T^{n-1} \quad (17)$$

where $a_P^T = \frac{(\rho c_p)_{i,j}^n}{\Delta t}r_P\Delta z\Delta r - \left(a_E^T+a_W^T+a_N^T+a_S^T\right)$,

$$a_E^T = - r_e k_e\frac{\Delta z}{\Delta r},\ a_W^T = - r_w k_w\frac{\Delta z}{\Delta r}$$

$a_N^T = - r_n k_n\frac{\Delta r}{\Delta z},\ a_S^T = - r_s k_s\frac{\Delta r}{\Delta z}$ and

$$S_T^{n-1}=\begin{bmatrix}\left(\frac{5k_B}{4e}r_P\Delta r\left(J_{z,n}^{n-1}+\left|J_{z,n}^{n-1}\right|\right)-\frac{r_n\Delta r}{2}\rho_n^{n-1}\left(u_n^{n-1}-\left|u_n^{n-1}\right|\right)\right)\left(T_{i+1,j}^{n-1}-T_{i,j}^{n-1}\right)\\+\left(\frac{5k_B}{4e}r_P\Delta r\left(J_{z,s}^{n-1}-\left|J_{z,s}^{n-1}\right|\right)-\frac{r_s\Delta r}{2}\rho_s^{n-1}\left(u_s^{n-1}+\left|u_s^{n-1}\right|\right)\right)\left(T_{i,j}^{n-1}-T_{i-1,j}^{n-1}\right)\\+\left(\frac{5k_B}{4e}r_P\Delta z\left(J_{r,e}^{n-1}+\left|J_{r,e}^{n-1}\right|\right)-\frac{r_e\Delta z}{2}\rho_e^{n-1}\left(v_e^{n-1}-\left|v_e^{n-1}\right|\right)\right)\left(T_{i,j+1}^{n-1}-T_{i,j}^{n-1}\right)\\+\left(\frac{5k_B}{4e}r_P\Delta z\left(J_{r,w}^{n-1}-\left|J_{r,w}^{n-1}\right|\right)-\frac{r_w\Delta z}{2}\rho_w^{n-1}\left(v_w^{n-1}+\left|v_w^{n-1}\right|\right)\right)\left(T_{i,j}^{n-1}-T_{i,j-1}^{n-1}\right)\end{bmatrix}+\left(\frac{\rho c_p T}{\Delta t}+\frac{J\cdot J}{\sigma}-S_R\right)_{i,j}^{n-1}\cdot r_P\Delta z\Delta r.$$

The discretized equation for $\varphi$:

$$a_S^{\varphi}\varphi_{i-1,j}^n+a_W^{\varphi}\varphi_{i,j-1}^n+a_P^{\varphi}\varphi_{i,j}^n+a_N^{\varphi}\varphi_{i+1,j}^n+a_E^{\varphi}\varphi_{i,j+1}^n=0 \qquad (18)$$

where $a_P^{\varphi}=-\left(a_E^{\varphi}+a_W^{\varphi}+a_N^{\varphi}+a_S^{\varphi}\right)$,

$$a_E^u=-r_e\sigma_e\frac{\Delta z}{\Delta r},\ a_W^u=-r_w\sigma_w\frac{\Delta z}{\Delta r}$$

$$a_N^u=-r_n\sigma_n\frac{\Delta r}{\Delta z}\ \text{and}\ a_S^u=-r_s\sigma_s\frac{\Delta r}{\Delta z}.$$

- SIMPLE algorithm

In FVM, the plasma density is generally not solved directly through the mass continuity equation. Instead, one needs to derive the algebraic equation for solving pressure according to the mass and momentum equations, and then to determine the density via the equation of state. This idea is the famous SIMPLE (Semi-Implicit Method for Pressure Linked Equations) algorithm [31] and is also used here.

If pseudo velocities are used, the final algebraic equations for solving $u$ and $v$ can be written as the following forms:

$$u_{i,j}^n=\hat{u}_{i,j}^n-d_{i,j}^u\left(p_n^n-p_s^n\right) \qquad (19)$$
$$v_{i,j}^n=\hat{v}_{i,j}^n-d_{i,j}^v\left(p_e^n-p_w^n\right) \qquad (20)$$

where $\hat{u}_{i,j}^n$ and $\hat{v}_{i,j}^n$ are pseudo velocities. $d_{i,j}^u$ and $d_{i,j}^v$ are coefficients in front of the pressure.

After integrating Eq. (1) (mass conservation equation) over the controlled volume represented by the node $P$, we have:

$$\frac{\rho_{i,j}^n-\rho_{i,j}^{n-1}}{\Delta t}r_P\Delta z\Delta r+(r\rho^n u^n)_n\Delta r-(r\rho^n u^n)_s\Delta r+(r\rho^n v^n)_e\Delta z-(r\rho^n\rho v^n)_w\Delta z=0 \qquad (21)$$

For the collocated grid shown in Fig. 2, we can make the velocities at boundaries of each controlled volume take the form similar to the velocities at nodes. Therefore, the velocities on the boundaries read:

$$v_e=\hat{v}_e-d_e\left(p_{i,j+1}-p_{i,j}\right) \qquad (22)$$

$$v_w=\hat{v}_w-d_w\left(p_{i,j}-p_{i,j-1}\right) \qquad (23)$$

$$u_n=\hat{u}_n-d_n\left(p_{i+1,j}-p_{i,j}\right) \qquad (24)$$

$$u_s=\hat{u}_s-d_s\left(p_{i,j}-p_{i-1,j}\right) \qquad (25)$$

where the super script $n$ is not specially labelled for convenience. $d_e$, $d_w$, $d_n$, $d_s$ and the pseudo velocities $\hat{v}_e$, $\hat{v}_w$, $\hat{u}_n$ and $\hat{u}_s$ are all determined by the linear interpolation from corresponding node values. For example,

$\hat{v}_e$ and $d_e$ are expressed as:

$$\hat{v}_e = \hat{v}_{i,j}\frac{(\delta r)_e^{+}}{(\delta r)_e} + \hat{v}_{i,j+1}\frac{(\delta r)_e^{-}}{(\delta r)_e}$$

$$d_e = d_{i,j}^{v}\frac{(\delta r)_e^{+}}{(\delta r)_e} + d_{i,j+1}^{v}\frac{(\delta r)_e^{-}}{(\delta r)_e}$$

The substitution of Eqs. (22)-(25) into the Eq. (21) finally gives the algebraic equation for pressure as the following:

$$a_E^p p_{i,j+1} + a_W^p p_{i,j-1} + a_P^p p_{i,j} + a_N^p p_{i+1,j} + a_S^p p_{i-1,j} = S_p \qquad (26)$$

where $a_P^p = -\left(a_E^p + a_W^p + a_N^p + a_S^p\right)$,

$$a_E^p = -\rho_e d_e r_e \Delta z,\ \ a_W^p = -\rho_w d_w r_w \Delta z$$

$$a_N^p = -\rho_n d_n r_n \Delta r,\ \ a_S^p = -\rho_s d_s r_s \Delta r, \text{ and}$$

$$S_p = \frac{\rho_{i,j}^{n} - \rho_{i,j}^{n-1}}{\Delta t} r_P \Delta z \Delta r + [(r\rho\hat{v})_w - (r\rho\hat{v})_e]\Delta z + [(r\rho\hat{u})_s - (r\rho\hat{u})_n]\Delta r.$$

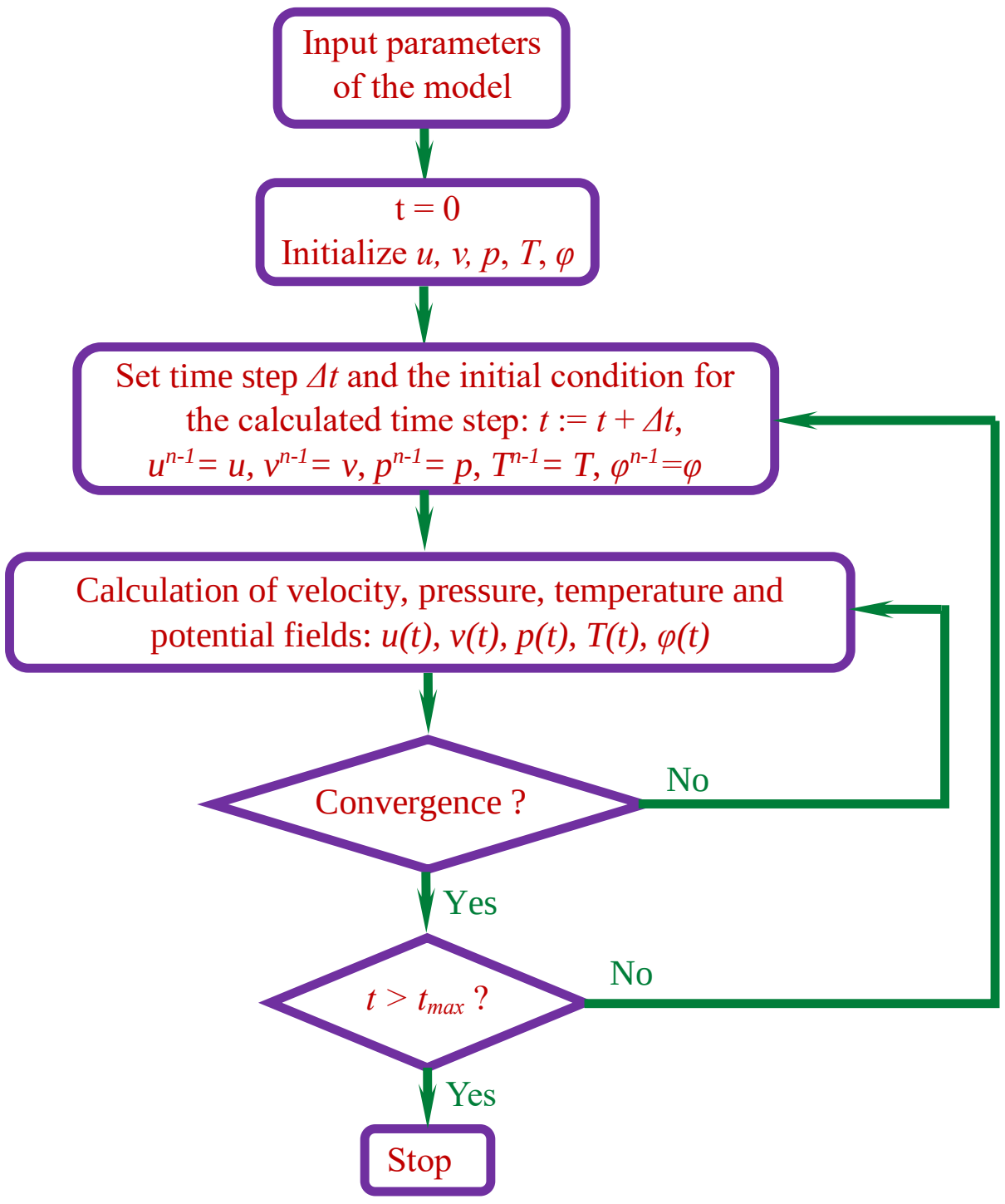

**Figure 3.** Integration scheme of simulations.

### 2.4 Numerical solution

Algebraic equations above can be solved by the iterative method. The following Gauss-Seidel iteration is used to accelerate calculations.

$$\phi_{ij}^{k+1} = a_W\phi_{i,j-1}^{k+1} + a_S\phi_{i-1,j}^{k+1} + a_E\phi_{i,j+1}^{k} + a_N\phi_{i+1,j}^{k} + b^k$$

where $k$ denotes the number of iterations.

To improve the convergence of discretized equations can employ the relaxation iteration, which takes the form:

$$\phi_{ij}^{k+1} = (1 - \alpha)\phi_{ij}^{k} + \alpha\phi_{ij}^{k+1}$$

where $\alpha$ is the relaxation factor and is in the range of 0~1.

Before the iteration process for each time step is finished, the following iteration criteria must be satisfied:

$$\left|\frac{p^{k+1} - p^k}{\alpha_p p^k + \varepsilon_0}\right| \leq \epsilon^p,\ \left|\frac{T^{k+1} - T^k}{\alpha_T T^k + \varepsilon_0}\right| \leq \epsilon^T,\ \left|\frac{\varphi^{k+1} - \varphi^k}{\alpha_\varphi \varphi^k + \varepsilon_0}\right| \leq \epsilon^\varphi$$

$$\left|\frac{u^{k+1} - u^k}{\alpha_u u^k + \varepsilon_0}\right| \leq \epsilon^u \text{ and } \left|\frac{v^{k+1} - v^k}{\alpha_v v^k + \varepsilon_0}\right| \leq \epsilon^v$$

where $\epsilon^p$, $\epsilon^T$, $\epsilon^\varphi$, $\epsilon^u$, and $\epsilon^v$ are iteration criteria for $p$, $T$, $\varphi$, $u$ and $v$, respectively, and they can be selected from $10^{-3}$~$10^{-6}$. $\alpha_p$, $\alpha_T$, $\alpha_\varphi$, $\alpha_u$ and $\alpha_v$ are under-relaxation coefficients for $p$, $T$, $\varphi$, $u$ and $v$, respectively. $\varepsilon_0$ is a very small number which is chosen to avert the data overflow.

The integration scheme in our simulation is given in Fig. 3.

## 3. Results and discussion

### 3.1 Model validation

In this section, the developed calculation model is implemented on the free-burning argon arc at the atmospheric pressure to verify its numerical accuracy. The arc current is 200 A, arc length is 10 mm, and the tip angle and the cut radius of cathode are respectively 45° and 0.33 mm. Thermal physical properties of argon including specific heat, viscosity, thermal and electrical conductivities are given in Fig. 4. Those data are referred from previous literatures [32-34].

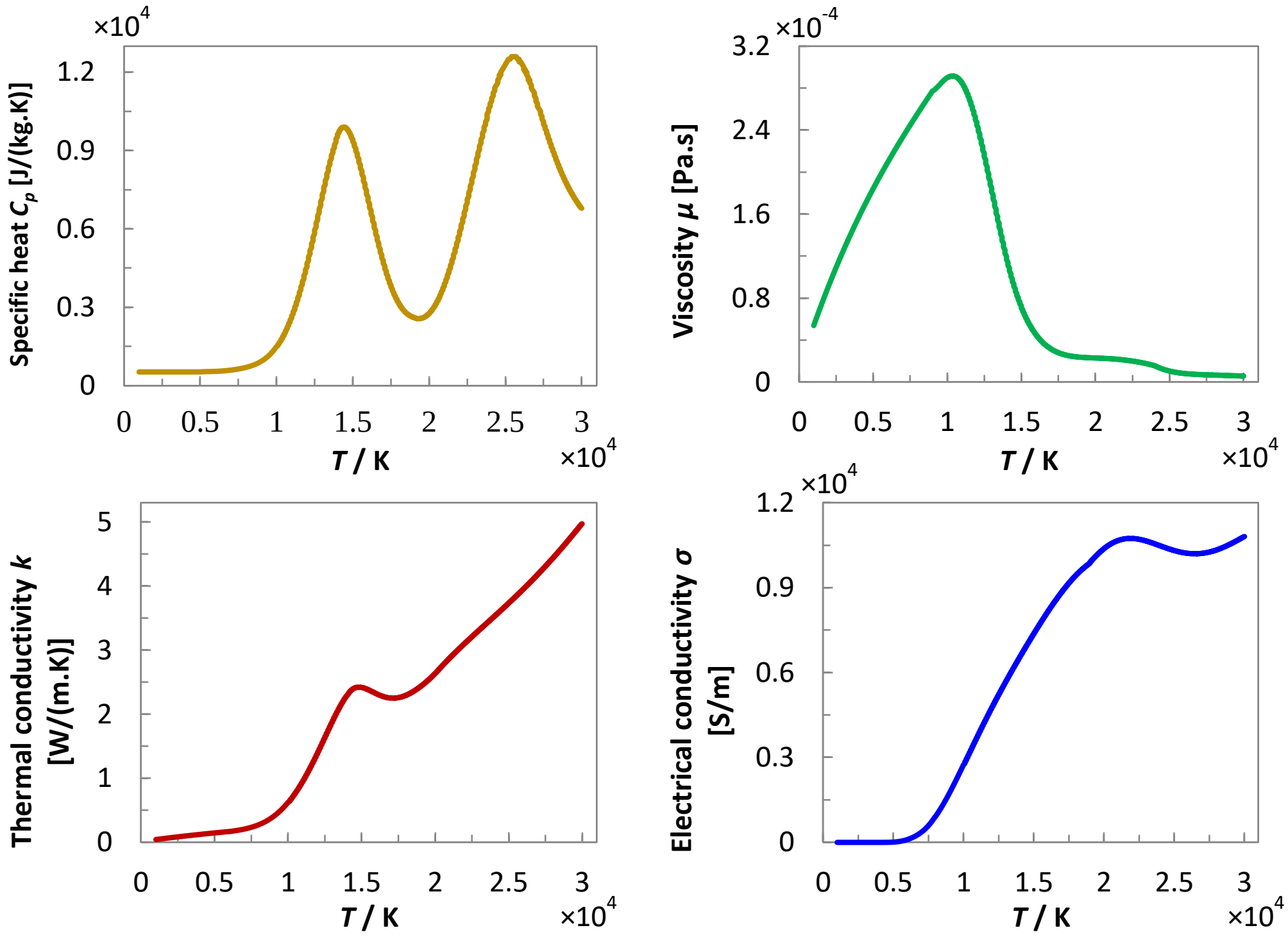

**Figure 4.** Thermal physical properties of argon at different temperatures.

We compare our calculation results with the available experimental data and numerical predictions [1] in Fig. 5 and Table 2. Fig. 5 shows the comparison in arc temperature field. Listed in Table 2 are key arc parameters, including the maximum temperature $T_{max}$, maximum axial velocity $U_{max}$, the overpressure at the cathode tip $P_{cathode}$ and the center of anode surface $P_{anode}$, axial current density at the center of anode surface $J_z^{anode}$ and the voltage drop between cathode and anode $\varphi_D$. From Fig. 5 and Table 2, we can see clearly that our numerical predictions are in good agreement with the experimental data and the calculations by Hsu et al. [1], demonstrating the enough numerical accuracy of our model. Especially, in Fig. 5 the well-known "bell" shape of a free-burning arc is confirmed again by our calculations.

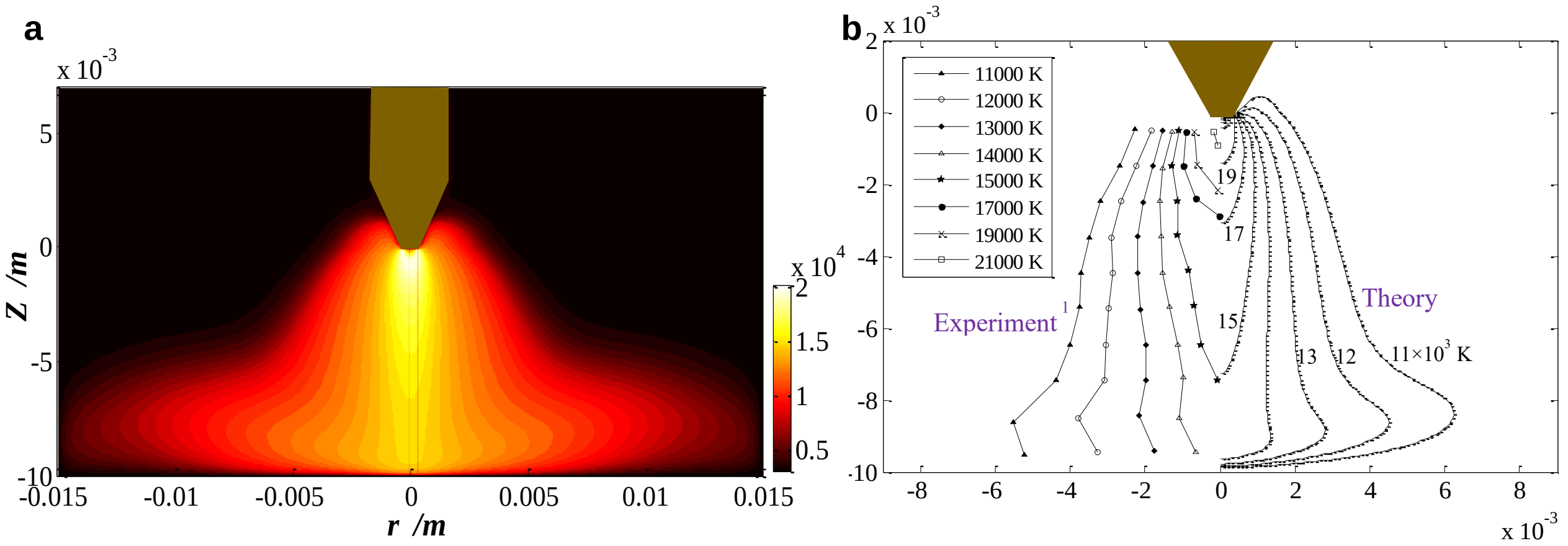

**Figure 5. Arc temperature field. a**) Computed arc temperature field $T$, K. **b**) Comparison with experimental data[1]. Arc current $I$ = 100 A, and arc length $L$ = 10 mm.

**Table 2: Comparison in key arc parameters.**

| Parameters | Hsu et al.[1] | Our results |
|---|---|---|
| $T_{max}$, K | 21200 | 20758 |
| $U_{max}$, m/s | 294 | 290 |
| $P_{cathode}$, Pa | 842 | 852 |
| $P_{anode}$, Pa | 394 | 470 |
| $J_z^{anode}$, A/m$^2$ | $3.1\times10^6$ | $2.9\times10^6$ |
| $\varphi_{cathode}$, V | 13.3 | 11.3 |

Note that there are two evident differences in values of $P_{anode}$ and $\varphi_{cathode}$. It is probably that in our simulations, the anode surface has been assumed to be at a fixed temperature of 3000 K. In reference [1], the temperature (enthalpy) distribution at the anode surface was provided by the experimental data, which are unknown to us. This different treatment of the boundary condition of anode surface may cause the difference in the predicted pressure distribution at anode surface. Besides, the whole cathode region is coupled to calculations in our simulation, but it was not considered by Hsu et al. [1]. This may cause the different computed values of $\varphi_D$. When we utilize the same boundary condition for the current density distribution at the cathode surface as Hsu et al. [1], the calculated voltage drop $\varphi_D$ is also very closed to 13.3 V. Furthermore, though we have adopted the boundary condition for the temperature distribution at the anode surface that is different from Hsu et al. [1], the computed value of $J_z^{anode}$ ($2.9\times10^6$ A/m$^2$) is still very close to that in [1] ($3.1\times10^6$ A/m$^2$), suggesting that our treatment of the temperature distribution at anode surface has little impact on the current density distribution.

Fig. 6 presents spatial distributions of the absolute velocity ($V=\sqrt{u^2+v^2}$), overpressure (relative to the atmospheric pressure), current density and the self-induced magnetic field intensity. It can be observed that the velocity field has a very sharp variation in the radial direction, and only near the arc's axis (except the locations in front of electrodes) the velocity $V$ is in the high level. The overpressure field is observed to exhibit a "tower" shape and the intensity of overpressure locally concentrates at the cathode tip and the center of anode surface. Besides, the current density has very high values (around 1.6×10$^8$ A/m$^2$) near the cathode tip. These high values have induced the initial triggering of arc. The self-induced magnetic field intensity $B_\theta$ has a peak (about 0.045 T) locating at the cathode surface which is above the tip. These observations are within our expectations, since argon gas only burns locally and intensively near the cathode tip.

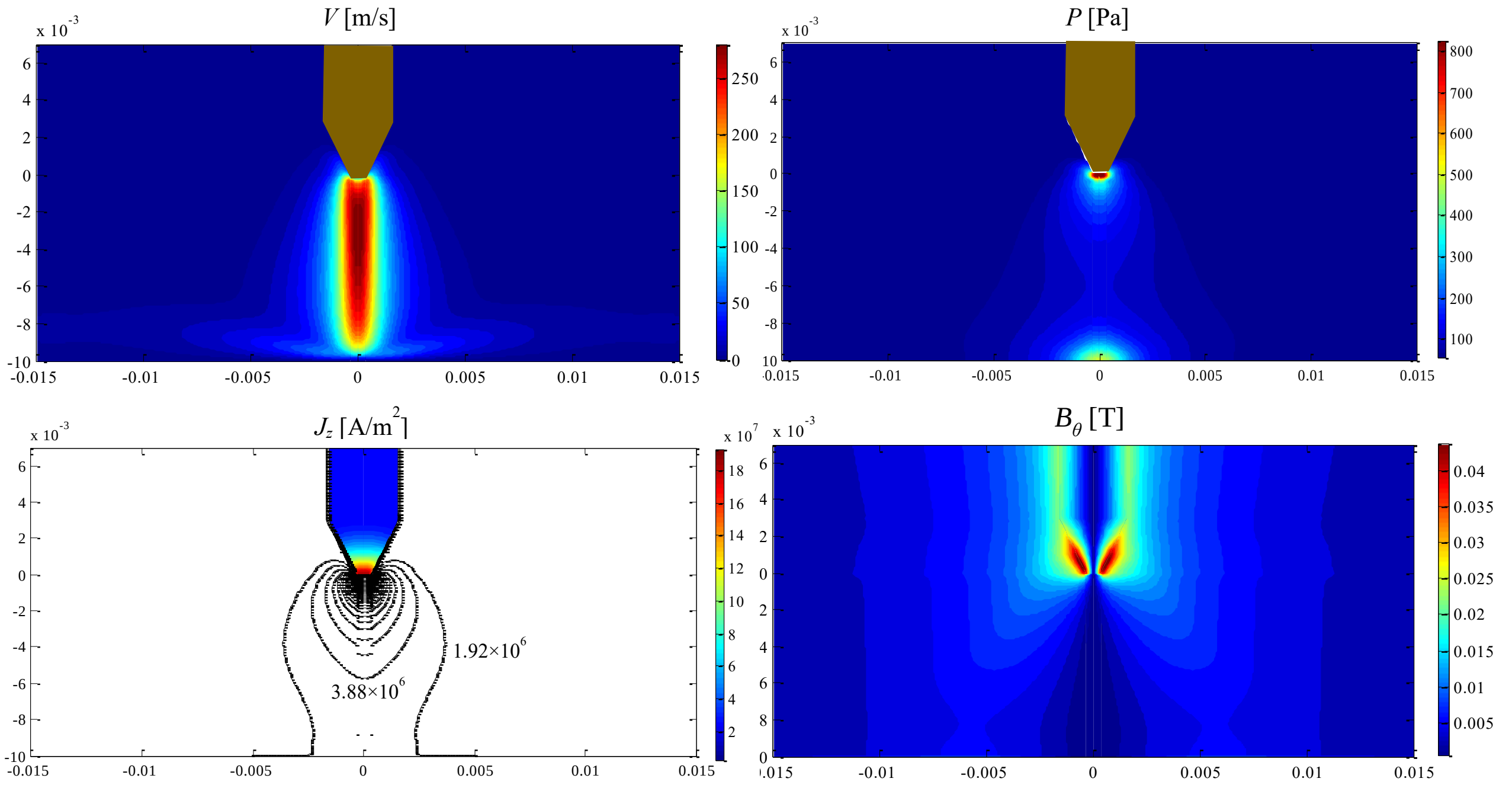

**Figure 6.** Spatial distributions of flow speed $V$, overpressure $P$, axial current density $J_z$ and toroidal magnetic field intensity $B_\theta$. Horizontal axis ($r$ coordinate) and longitudinal axis ($z$ coordinate) are in m.

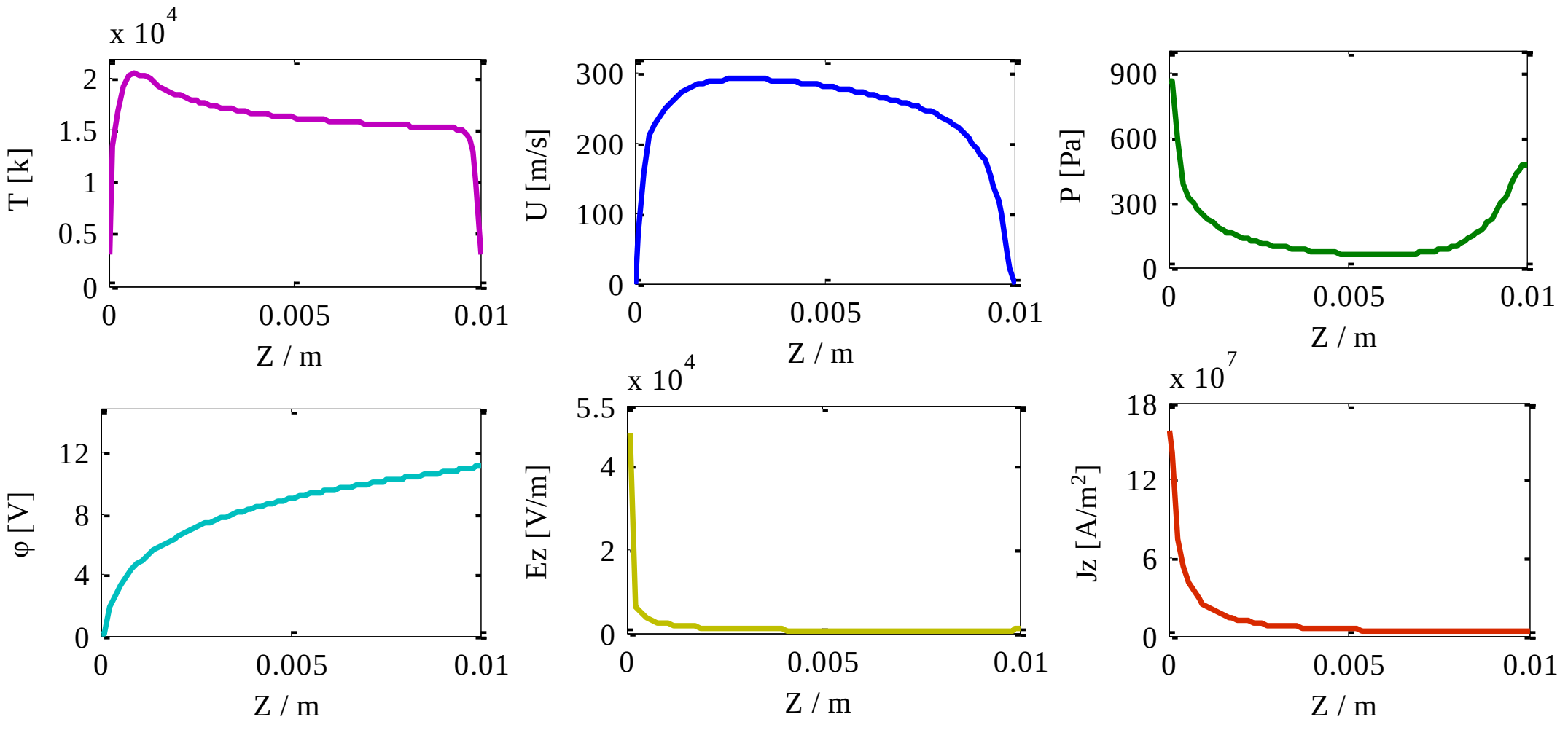

**Figure 7.** Variation of temperature $T$, axial flow velocity $U$, overpressure $P$, electrical potential $\varphi$, field strength $E_z$, and axial current density $J_z$ on the axis of the free-burning argon arc.

In Fig. 7, we plot the centerline arc temperature, the axial velocity component, the overpressure, electrical potential, electric field strength and the axial current density. Consistent with the reference [1], the temperature, axial velocity and overpressure all rapidly vary in front of both cathode tip and anode surface, and the electrical potential, electric field strength and the axial current density only sharply increase or decrease in front of the cathode tip.

The shear stress, generated by the sweep of plasma over the anode surface, results in a transfer of momentum from the plasma to anode. In practical applications, the stress will affect the fluid flow in the weld pool and the subsequent structure of the weld, and should be known. According to the Newton's law of inner friction, the shear stress can be defined as follows:

$$\tau_{anode} = \left(\mu \frac{dv}{dz}\right)\Big|_{anode} \qquad (27)$$

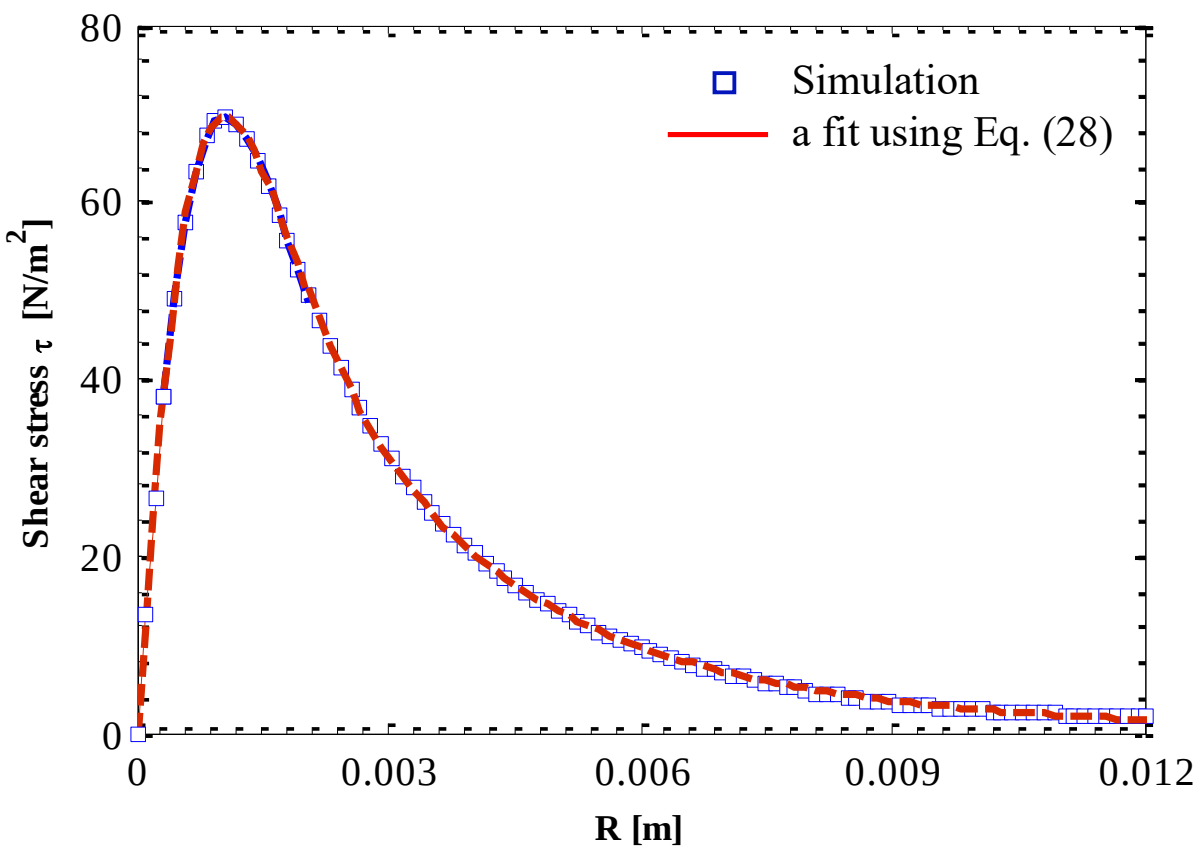

**Figure 8.** Radial distribution of the shear stress at the anode surface.

As shown in Fig. 8, the shear stress distributed on the anode surface has a peak (about 70 N/$m^2$) around $r$ = 1 mm. Besides, at both the center of anode surface and the location far from arc plasma region, the stress reduces to zero. Specially, the shear stress is observed to have a tail ($r$ > 2 mm) which decays as an simple exponential law. The following mathematical function is proposed to describe the whole radial distribution of shear stress.

$$\tau_{anode}\ (\mathrm{N/m^2}) = \begin{cases} r^{\alpha} e^{-\beta r + c}, & 0 < r < 2\ mm \\ c_1 + e^{-\gamma r + c_2}, & r > 2\ mm \end{cases} \qquad (28)$$

where $\alpha = 1.2567$, $\beta = 1192.73$, $c = 14.11$, $c_1 = 1.32$, $\gamma = 453.05$, and $c_2 = 4.78$. $r$ is in m. This function has been included in the figure and is observed to well fit data.

### 3.2 Constricting arc with alternating magnetic field

As mentioned before, to constrict arc plasma is of interest to the welding field. Here, we report an unexpected observation that the applied high-frequency alternating longitudinal magnetic field is able to make argon arc shrink intensively. In this case, the local hollow region near the anode, which tends to appear in the constant axial magnetic field, will disappear, and the confinement produced on the arc plasma will also become more effective, compared with the constant magnetic field (Fig. 9). In our simulations, we find that the strongly shrinked arc is not in a still state but in a dynamic state which continuously switches between shrinking and expansion, and the applied alternating magnetic field can play a "plasma trap" role, which succeeds to imprison the arc plasma into a much narrower space. This indicates that the dynamic confinement on arc plasma, to some extent, seems better.

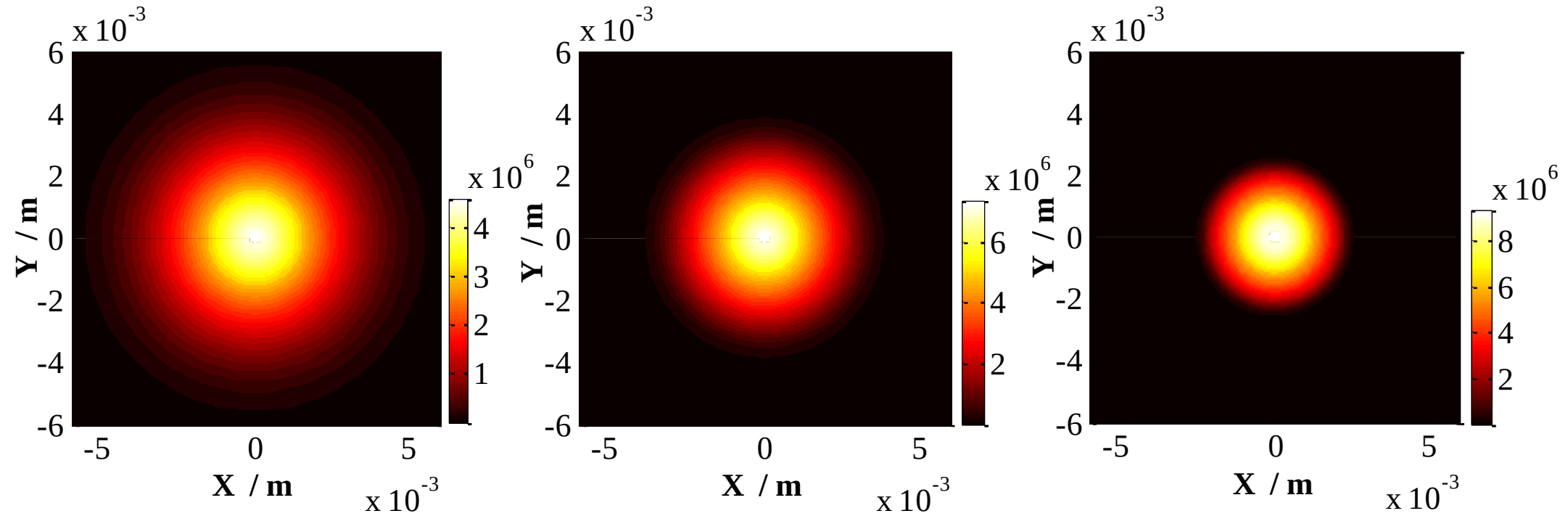

**Figure 9.** Period-averaged spatial distribution of current density at the cross section of Z = -3 mm, A/m$^2$. **a**): free arc; **b**): $B_{0z}$ = 30 mT and $f_m$ = 0 Hz. c): $B_{0z}$ = 30 mT and $f_m$ = 1.5 kHz. Arc current $I$ = 100 A and arc length $L$ = 5 mm.

To disclose the mechanism behind the above phenomenon, we need to analyze the motion of arc plasma. In our simulations, the equation describing plasma's motion is:

$$\rho\frac{d\boldsymbol{V}}{dt} = -\nabla p + \nabla\cdot\left(2\mu\widehat{\boldsymbol{S}}\right) + \boldsymbol{J}\times\boldsymbol{B} \quad (27)$$

Where $\boldsymbol{V} = (v_z, v_r, v_\theta)$ is the velocity vector of arc plasma, $p$ is the thermal pressure, $\widehat{\boldsymbol{S}}$ is the velocity's deformation rate tensor, $\boldsymbol{J} = (J_z, J_r, 0)$ is the arc current density, and $\mu$ is the viscosity. $\boldsymbol{B} = (B_{0z}, 0, B_\theta)$ is the magnetic field vector, where $B_\theta$ is the arc's self-magnetic field induced by the arc current density $J_z$, and $B_{0z}$ is the applied high-frequency alternating axial magnetic field (see Fig. 12).

Eq. (27) has assumed that the arc flow is laminar, and the weak toroidal current produced by arc's rotation is negligible. Eq. (27) suggests that the forces arc plasma mainly sustains during its motion mainly include the pressure, viscous force and Lorentz force.

Within the real arc, the motion path along which a small cluster of plasma runs from the cathode to anode is generally a complicated curve and there is no force balance in axial or radial directions, even for the simplest free-burning arc (see Fig. 5). To simplify the analysis and also without the loss of generality, we can analyze a simpler arc plasma system that the whole arc plasma region is cylindrical shaped and is infinitely long so that the arc property in each axial plane is similar. When this system is under the constant axial magnetic field, a small cluster of plasma with a mass of $m_e$ and volume $V_e$ will do the helical motion at constant speeds of $v_\theta$ and $v_z$ and the radius of $R_0$. The centripetal force for the circular motion $F_c = m_e v_\theta^2/R_0$ is mainly provided by the sum $F_r$ of the radial pressure and the Lorentz force $F_{Br} = -J_z B_\theta$ (always in $r_-$ direction). In the circumferential direction, the Lorentz force $F_{Bt} = -J_r B_{0z}$, which is produced by the applied magnetic field and induces arc plasma to rotate, is balanced with the viscous resistance $F_{\mu t}$ induced by the velocity shear.

If at one point the applied longitudinal magnetic field is in reverse direction, $F_{Bt}$ will also be in the opposite direction immediately and become $F_{Bt}'$, and then work together with the viscous resistance $F_{\mu t}$ to drag this small cluster to slow down its rotation (Fig. 10a). During the slowing down of rotation, this cluster will be gradually hauled to the lower orbit by the relatively stronger sucking force (the radial force $F_r$). This process seems very similar to the well-known phenomenon that artificial satellites always fall down under the earth gravity once their speed slows down due to some factors. In the process that the cluster rotates inward, the Lorentz force $F_{Br}$, which drives the cluster to move inward, will further increase since $J_z$ and $B_\theta$ will be strengthened according to the current conservation. This additional effect will further drive the cluster to move towards the arc's axis. Under the action of the reverse circumferential Lorentz force $F_{Bt}'$, however, after reducing to zero the rotation speed of the cluster will increase gradually in the opposite direction (see Fig. 12). Meanwhile, the inward radial force $F_r$ will decrease slowly and then increase in the outward direction ($r_+$ direction). Therefore, after moving inward a specific distance, this minor cluster will rotate outward. The rotation speed of plasma in the alternating magnetic field is smaller than in the constant magnetic field on the time-weighted average, because the alternating magnetic field causes arc's rotation to undergo extra slowing down processes. Thereby, this cluster cannot return to the original orbit that is under the constant magnetic field. After several rounds of above repetitive process, this small cluster of plasma will finally do the continuous inward and outward rotation motion, within a narrower annular space of the inner radius $R_2$ and the outer radius $R_1$. The cluster of plasma seems to be imprisoned into an annular trap and cannot escape from it any more.

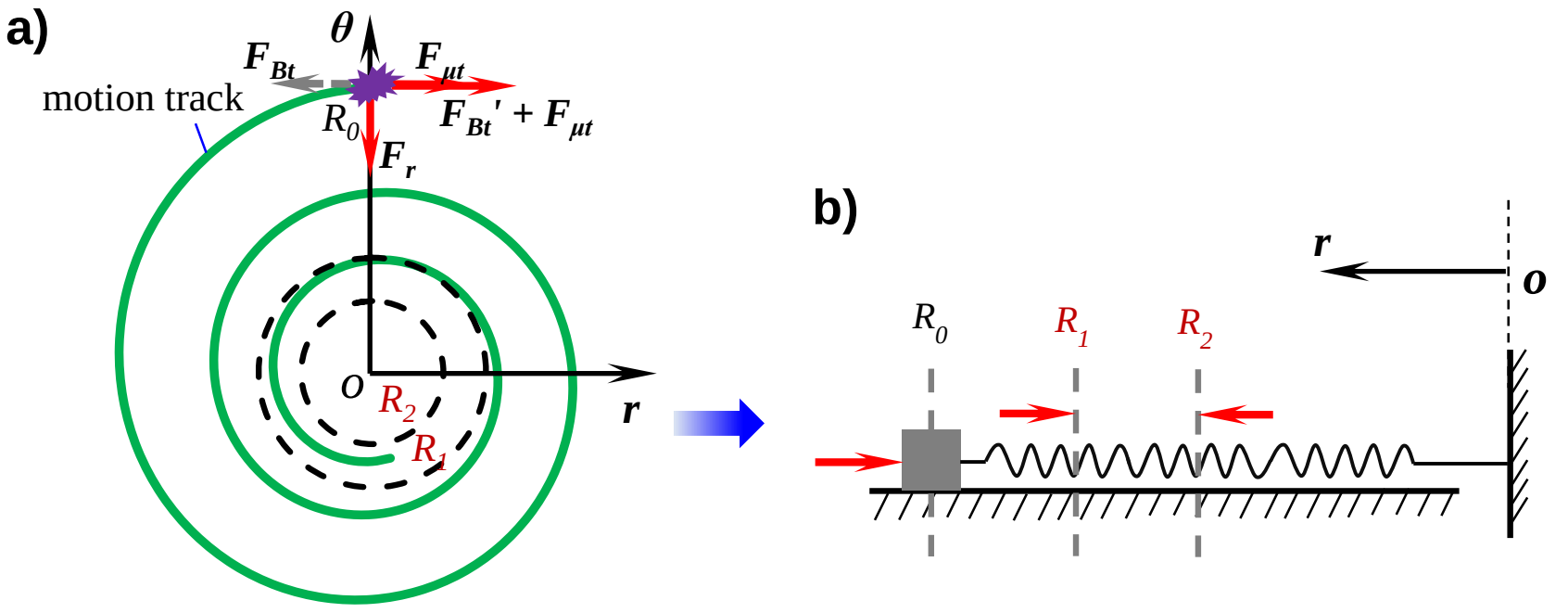

**Figure 10.** Illustration of the motion of arc plasma in the alternating longitudinal magnetic field.

The back-and-forth motion in the radial direction in Fig. 10a can be further abstracted to the motion of a spring oscillator in Fig. 10b. When the cluster is at $R_1$, its velocity is zero, but it sustains the largest inward pulling force. So, it will then move in the opposite direction until reaching $R_2$. At $R_2$, the velocity reduces to zero again, but at this time it sustains the largest outward pushing force. In this way, the cluster repeatedly moves between locations $R_1$ and $R_2$. Of course, within a real arc, plasma's motion will be more complicated, but the general process is similar. We call the role played by the applied high-frequency alternating longitudinal magnetic field in arc the "plasma trap", which can effectively pinch the arc plasma. Some similar concepts using the proper magnetic field to confine charged particles have already been put forward and applied, like the famous "Paul Trap" [35], which has been applied to the long-distance confinement of charged particle beam in the accelerator.

Note that in the applied alternating axial magnetic field, it is basically the arc's inertia nature that is at work and causes the arc to further shrink. One can imagine that if all arc parameters (e.g. the rotation speed) finish their changes instantly as the applied magnetic field does, only the rotation direction of plasmas will become opposite, which will hardly make arc to shrink. The alternating magnetic field actually provides the proper chance for plasma's motion inertia to play its role. This can be proved by the results shown in Fig. 12, where the deceleration/relaxation time (about 0.2 ms) of arc's rotation speed is much extended, relative to the zero time that the magnetic field takes to change its direction.

Fig. 11 shows the relaxation process of TIG arc in the alternating longitudinal magnetic field. Initially ($t$ = 0), arc is in the constant axial magnetic field (Fig. 11a). At this time, the alternating magnetic field with a frequency of 1.5 kHz is imposed, and its direction becomes opposite when $t$ = 0.333 ms. It can be observed that the first arc shrinking (Fig. 11b) occurs at 0.4 ms evidently. After undergoing several shrinking and expansion cycles, the arc finally reaches a stable dynamic state and continuously shrinks and expands between the two states shown in Figs. 11c, d. In the meantime, the local hollow region, which tends to occur near the anode in the constant axial magnetic field, is also observed to disappear. These results indicate that in the alternating magnetic field, the arc achieves to be pinched through the continuous dynamic transition between shrinking and expansion. The change in the spatial distribution of the arc current density in the axial plane also shows the more effective confinement of alternating magnetic field on the arc plasma, compared with the constant magnetic field (Fig. 9).

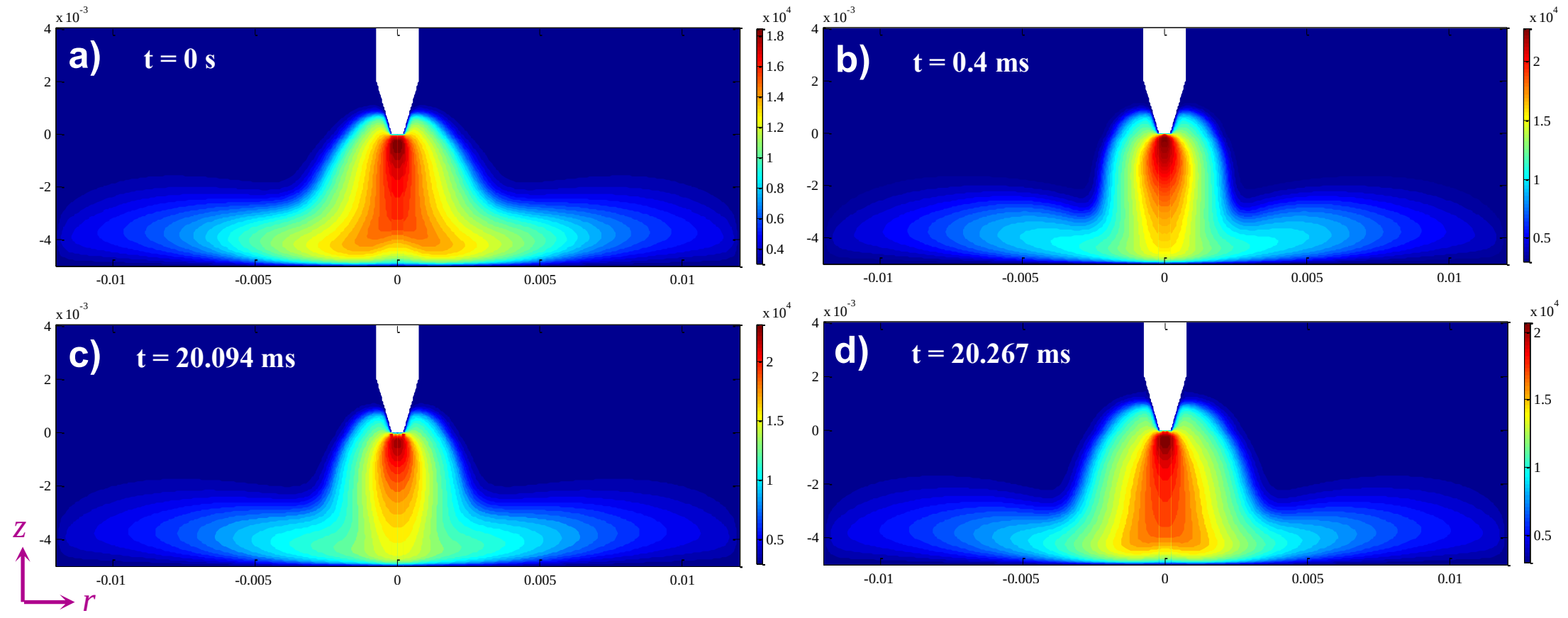

**Figure 11.** Relaxation processes of arc in the alternating axial magnetic field (temperature distribution, K). Arc current $I$ = 100 A, arc length $L$ = 5 mm, $B_{0z}$ = 30 mT, $f_m$ = 1.5 kHz, and time step length $t_p \approx 2\times10^{-7}$ s.

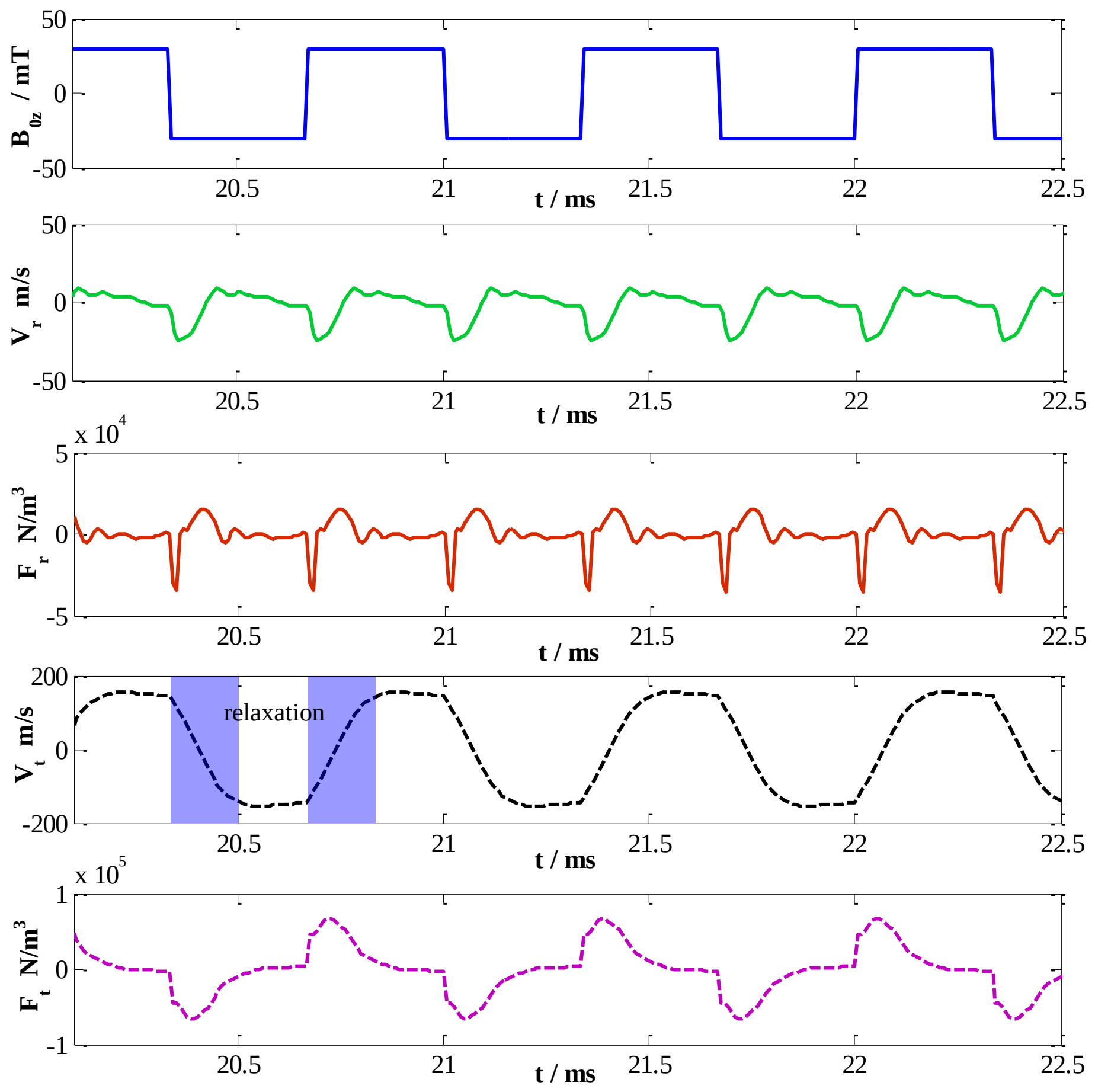

**Figure 12.** Evolution of radial velocity $v_r$, resultant radial force $F_r$, toroidal velocity $v_t$ and the toroidal force $F_t$ at the position of $Z$ = -2 mm and $r$ = 1 mm. $B_{0z}$ = 30 mT and $f_m$ = 1.5 kHz.

Fig. 12 plots evolutions of some key arc parameters at one fixed location after the arc reaches its stable state in the applied magnetic field. Evolutions of these parameters are observed to be generally consistent with previous analysis. In addition, the stable periodic evolutions of these parameters suggest that the final arc is stably in a rapid radial oscillation state.

Note that under the alternating magnetic field, the final arc still shrinks and expands continuously and has no fixed geometric configuration. The arc current density given in Fig. 9 is the time-average result within one magnetic field period.

In our simulations, we also observed that for one specific magnetic field intensity, there exists one optimal frequency $f_{op}$ that can pinch the arc plasma most effectively ($f_{op} \approx$ 1.5 kHz when $B_{0z}$ = 30 mT). The reason may be that the arc has its own eigen frequency, when the applied magnetic field frequency is close to this value, then the arc is more likely to interact with the external magnetic field and get better confined. It is also observed that the confinement of high-frequency alternating axial magnetic field on arc plasma is effective within the range $B_{0z}$ = 10 ~ 100 mT.

## 4. Summary

The detailed pressure-based finite volume simulation of arc is presented. The model is validated with experiment in case of the free-burning argon arc under the atmospheric pressure. The shear stress on the anode surface is observed to have a peak around $r$ = 1 mm and an exponentially decaying tail ($r$ > 2 mm). We observe an interesting phenomenon that arc can be constricted by the applied high-frequency alternating longitudinal magnetic field. The final arc is in the relaxation dynamics which continuously switches between shrinking and expansion, and the confinement produced by the alternating magnetic field is more effective than the constant magnetic field. The behind mechanism is that the applied high-frequency alternating magnetic field is able to cooperate with plasma's motion inertia to effectively play the "plasma trap" role, which imprisons the arc plasma into a narrower space. Our result suggests that the dynamic confinement, to some extent, is better. This finding not only helps to get a deeper insight into behaviors of arc, but also provides a potential approach to confine arc plasmas.

**Acknowledgements:** This work is supported by University of Science and Technology of China and National Natural Science Foundation of China.

**Conflict of interest**: The author declares no conflict of interest.

**Data and materials availability**: All calculation data are available from X.W. upon reasonable request.